# The Complex History of Trojan Asteroids

**Joshua P. Emery**
***University of Tennessee***

**Francesco Marzari**
***Università di Padova***

**Alessandro Morbidelli**
***Observatoire de la Côte d'Azur***

**Linda M. French**
***Illinois Wesleyan University***

**Tommy Grav**
***Planetary Science Institute***

## Abstract

The Trojan asteroids, orbiting the Sun in Jupiter's stable Lagrange points, provide a unique perspective on the history of Solar System. As a large population of small bodies, they record important gravitational interactions and dynamical evolution of the Solar System. As primitive bodies, their compositions and physical properties provide windows into the conditions in the solar nebula in the region in which they formed. In the past decade, significant advances have been made in understanding physical properties, and there has been a revolution in thinking about the origin of Trojans. The ice and organics generally presumed to be a significant part of Trojan compositions have yet to be detected directly, though low density of the binary system Patroclus (and possibly low density of the binary/moonlet system Hektor) is consistent with an interior ice component. By contrast, fine-grained silicates that appear to be similar to cometary silicates in composition have been detected, and a color bimodality may indicate distinct compositional groups among the Trojans. Whereas Trojans had traditionally been thought to have formed near 5 AU, a new paradigm has developed in which the Trojans formed in the proto-Kuiper Belt, and they were scattered inward and captured in the Trojan swarms as a result of resonant interactions of the giant planets. Whereas the orbital and population distributions of current Trojans are consistent with this origin scenario, there are significant differences between current physical properties of Trojans and those of Kuiper Belt objects. These differences may be indicative of surface modification due to the inward migration of objects that became the Trojans, but understanding of appropriate modification mechanisms is poor and would benefit from additional laboratory studies. Many open questions about this intriguing population remain, and the future promises significant strides in our understanding of Trojans. The time is ripe for a spacecraft mission to the Trojans, to turn these objects into geologic worlds that can be studied in detail to unravel their complex history.

## 1. Introduction

Originally considered as simply an extension of the Main Belt, Trojan asteroids have become recognized as a large and important population of small bodies. Trojans share Jupiter's orbit around the Sun, residing the $L_4$ and $L_5$ stable Lagrange regions. Leading and trailing Jupiter by 60$^o$, these are regions of stable equilibrium in the Sun-Jupiter-asteroid three-body gravitational system. The moniker "Trojan" is an artifact of history – the first three objects discovered in Jupiter's Lagrange regions were named after heroes from the *Iliad*. The naming convention stuck for Jupiter's swarms, and the term Trojan eventually came to be used for any object trapped in the $L_4$ or $L_5$ region of any body. Nevertheless, only Jupiter Trojans are named from the *Iliad*, and when used without a designator, "Trojan" refers either specifically to Jupiter Trojans or sometimes to the collection of all bodies in stable Lagrange points. Several other Solar System bodies also support stable Trojan populations, including Mars, Neptune, and two satellites of Saturn (Tethys and Dione). The populations co-orbiting with Mars and the two saturnian moons appear to be quite small, but Neptune's family of Trojans is thought to be extensive (e.g., Sheppard and Trujillo 2010). Planets can destabilize each other's Lagrange regions. For instances, Saturn and Uranus do not have stable Trojan populations because the other planets perturb the orbits on timescales that are short relative to the age of the Solar System. The Jupiter Trojans, which are the focus of this chapter, are estimated to be nearly as populous as the Main Belt and have stability timescales that exceed the age of the Solar System.

The history of the exploration of Trojan asteroids begins with Max Wolf, who, in the late nineteenth century was the first to turn to wide-field astrophotography for asteroid discovery (Tenn 1994). In early 1906 he detected an object near Jupiter's $L_4$ point, marking the first observational confirmation of Lagrange's three-body solution. An object was detected near $L_5$ later in 1906 by August Kopf, then another near $L_4$ in early 1907. These were later named Achilles, Patroclus, and Hektor, respectively (Nicholson 1961). As physical studies of asteroids accelerated in the 1970s and 1980s, the Trojans were included, and the first sizes, albedos, rotation periods, and (visible wavelength) spectra were published (e.g., Dunlap and Gehrels 1969, Cruikshank 1977, Hartmann and Cruikshank 1978, Chapman and Gaffey 1979). Gradie and Veverka (1980) established the paradigm, which is still commonly invoked, that the low albedo and red spectral slopes are due to the presence of complex organic molecules on Trojan surfaces. By 1989, when the *Asteroids II* book was published, 157 Trojans were known, from which Shoemaker et al. (1989) estimated a total population comparable to that of the Main Belt – an estimates that still stands, to within a factor of a few. Discovery and characterization accelerated rapidly for Trojans (as with all asteroids) through the end of the twentieth century – by the time of *Asteroids III* in 2002, 993 Trojans had been discovered. The number now stands at 6073.

Summarizing the state of knowledge of the physical properties of Trojans at the turn of the twenty-first century, Barucci et al. (2002) describe a population that is far more homogeneous than the Main Belt, with uniformly low albedos ($p_v$ ~0.03 to 0.07) and featureless, red-sloped spectra at visible and near-infrared wavelengths (0.4 to 2.5 μm). A later review by Dotto et al. (2008) report additional spectral observations, particularly of members of potential collisional families (Dotto et al. 2006, Fornasier et al. 2007), the detection of signatures of fine-grained silicates (Emery et al. 2006), and the first bulk-density measurement (Marchis et al. 2006). From these properties and their locations at 5.2 AU, Trojans have generally been inferred to contain a large fraction of $H_2O$ ice, though hidden from view by a refractory mantle, and a higher abundance of complex organic molecules than most Main Belt asteroids. Since those reviews, significant strides have been made in the physical characterization of Trojans, which in turn provide new insights into the nature of these enigmatic bodies.

Marzari et al. (2002) review models for the capture of Trojans and the stability of the Lagrange regions that had developed up to that point. Although some analytical work suggested stability regions that did not match observations, numerical work by Levison et al. (1997) showed a wide region of stability for the age of the Solar System. Efforts to explain the capture of Trojans settled on two potential mechanisms as most likely: gas drag in the early nebula (e.g., Peale 1993) and capture during the growth of Jupiter (Marzari and Scholl 1998a). Both mechanisms predict that the present-day Trojans formed in the middle of the solar nebula, near where they currently reside. Since there is no other reservoir of material available for study from this region, the Trojans would, in this case, be an exciting window into the conditions of the solar nebula near the snow-line and near Jupiter's formation region. However, neither mechanism fully explains the current orbital properties of Trojans, particularly the high inclinations. More recently, Morbidelli et al. (2005) proposed the capture of Trojans from the same population from which the Kuiper Belt originated. The Nice model postulates that resonant interactions between Jupiter and Saturn temporarily destabilize the orbits of Uranus and Neptune, which move into the primordial Kuiper Belt, scattering material widely across the Solar System. In this framework, Jupiter's primordial Trojan population is lost and the Lagrange regions are repopulated with this scattered Kuiper Belt material. Dotto et al. (2008) include a description of this capture scenario and a discussion of the implications for Trojans. This mechanism predicts that Trojans formed much farther out in the solar nebula (~20 to 35 AU). In this case, the Trojans would represent the most readily accessible repository of Kuiper Belt material. In the years since those reviews, some aspects of the Nice model have been reworked, and refinements to this newer mechanism for Trojan capture have been made.

Unraveling the complex history of the Trojans promises key insight into Solar System evolution. As primitive objects, Trojan compositions provide direct indicators of the conditions of the nebula in the region(s) in which they formed. As a population of small bodies, Trojans act as unique probes of the history, interaction, and physical processing of the Solar System. In this chapter, we review the physical properties of Trojan asteroids and scenarios for their origin and evolution. We rely heavily on previous reviews for much of the early work (Shoemaker et al. 1989, Barucci et al. 2002, Marzari et al. 2002, Dotto et al. 2008), focusing here on new observations and recent advances in the knowledge of Trojans.

## 2. Physical Properties

### 2.1. Size Distribution

Most asteroid surveys are conducted in visible (reflected) light, from which it is not possible to derive the size unless the albedo is known. Studies of size distributions, therefore, often use absolute magnitude ($H_v$) as a proxy for size. For a population like the Trojans, where the albedo distribution is very uniform (see section 2.2), the $H_v$ distribution should closely match the actual size distribution. Shoemaker et al. (1989) pointed out that the largest Trojans have a fairly steeply sloped cumulative $H_v$ distribution (see Fig 1). Jewitt et al. (2000) measured the $H_v$ distribution for smaller Trojans, finding a shallower slope for their sample of objects with $H_v \gtrsim$ 10.5. They interpreted this as a break between a primordial population at larger sizes and a collisionally evolved population at smaller sizes. Trojan discoveries are now complete to a bit fainter than this ($H_V$~11.5, Karlsson 2010), so it is now possible to see the break in slope clearly at $H_V$~ 9 by downloading and plotting data from the IAU Minor Planet Center. Yoshida and Nakamura (2005) conducted their own survey of small Trojans in $L_4$. They confirmed the shallower slope for small Trojans found by Jewitt et al. and found a second break in the $H_V$-distribution at $H_V$~16 (D~5 km). In a follow-up study, Yoshida and Nakamura (2008) measured a similar slope in $L_5$ as the intermediate ($9 \lesssim H_V \lesssim 16$) size range for $L_4$, but no break for the smallest sizes. From the same survey, Nakamura and Yoshida (2008) confirm the previously recognized population asymmetry between the two swarms.

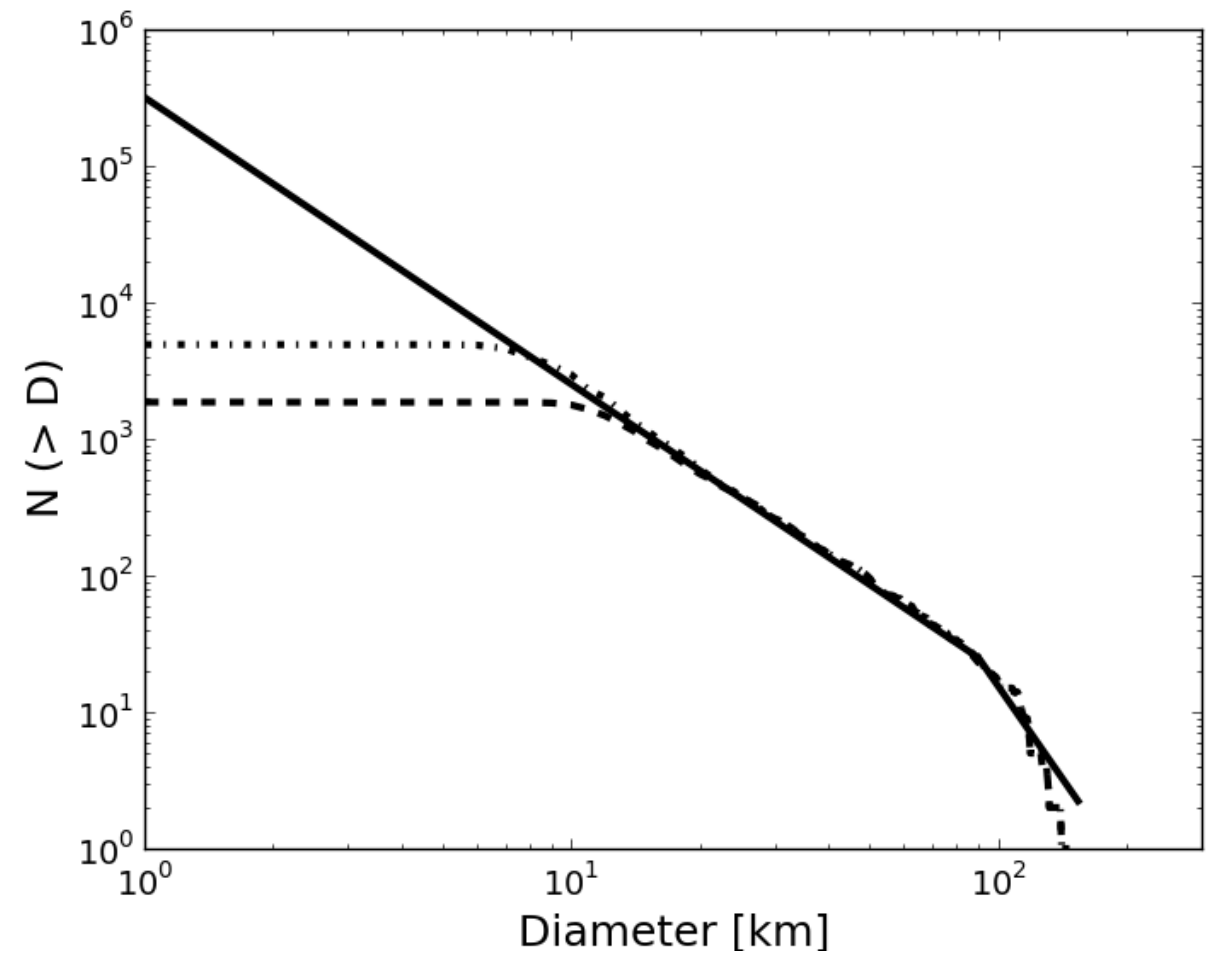

***Figure 1.*** *The size distribution of the Jovian Trojan population from the NEOWISE sample (dashed line) is complete for diameter larger than 20 km, while optical surveys currently sample to below 10 km (dotted line). Both samples show a size distribution with a cumulative power law index of ~2 (solid line) for diameters between 20 and 80 km. Modified from Grav et al. (2011).*

The NEOWISE project, an all sky infrared survey (Grav et al., 2011), allows for the direct derivation of diameters, since the Trojans were detected in thermal emission. This alleviates any uncertainty in converting from absolute magnitude to diameter. Figure 1 shows the cumulative size distribution of the NEOWISE sample compared to the known population of Jovian Trojans. The diameters of known objects that are not in the NEOWISE sample have been estimated using an albedo of 7% (average Trojan albedo from NEOWISE, see section 2.2) and their published absolute magnitudes from the Minor Planet Center catalog. The NEOWISE sample is nearly complete for diameters larger than ~20 km, whereas the known sample dominated by optical discoveries effectively reaches to objects with diameters of about 8-10 km. Grav et al. (2011) performed preliminary debiasing that showed a size distribution that is consistent with a power law of the form $N(>D) \sim D^{-a}$, where the power law index a = 2, when looking at the sample with diameters from 10 to 100 km. This is consistent with the earlier estimates by Jewitt et al. (2000), which investigated the size distribution of the smaller Jovian Trojans. They surveyed a 20 sq. degree field in the L4 cloud reaching a limiting magnitude of V = 22.5 and detected 93 Jovian Trojans with diameters from 4 to 40 km (where they assumed a visual albedo of 4%). They derived a power-law index of 2.0±0.3 for the absolute magnitude distribution in this size range.

At diameters larger than ~80 km the distribution is significantly steeper, but the sample is limited with only 34 objects. Shoemaker et al. (1989) estimated the cumulative power law index to 4.5 ± 0.9 for objects with diameters larger than 84 km. The NEOWISE results show that the slope is even steeper than that for the largest Jovian Trojans. Fraser et al. (2014) re-examined the absolute magnitude distribution of bright Trojans to compare it with that of Kuiper belt objects. They fit the observed H-distribution with an exponential law $N(H)=10^{\alpha H}$ in different H-ranges. The best-fit for bright objects was found to be $\alpha_1 = 1.0\pm0.2$, similar to the value found by Jewitt et al. (2000) and consistent with the size distribution exponent estimated in Shoemaker et al. (1989) and Grav et al. (2011). This fit is valid only up to a (r'-band) magnitude $H_{Break} = 8.4^{+0.2}_{-0.1}$ beyond which the exponent changes to $\alpha_2 = 0.36 \pm 0.01$, compatible with the slope found by Yoshida & Nakamura (2008) and the slope of the size distribution found by the NEOWISE survey.

Comparing the size distributions of Trojans and KBO is very important in order to test the hypothesis that Trojans are KBOs captured during the phase of giant planet dynamical instability (see section 4.1). Fraser et al. took into account that the albedos of Trojans and KBOs are different on average and that the albedos of red and blue KBOs are different from each other as well. In summary, they found that the parent populations of the hot classical KBOs and Trojans are statistically indistinguishable. Given that the Trojan and hot classical size distributions are distinct from other analog populations (Main Belt asteroids and cold classical KBOs, respectively) Fraser et al. conclude that Trojan asteroids are derived from the hot classical Kuiper Belt. The same comparison between Trojans and cold classical KBOs revealed that there is less than 1 in 1000 probability that those two populations are drawn from the same parent distribution. This is driven by the much steeper large object slope of the cold Kuiper belt magnitude distribution (with $\alpha_1$ = 1.6±0:3). According to models of dynamical capture of Trojans in the context of the Nice model (Morbidelli et al. 2005, Nesvorny et al. 2013), the bodies that are captured originate from whatever portion of the original Kuiper Belt is scattered. It is now considered most likely that the hot classical population is the relic of the planetesimal disk that was scattered all over the Solar System at the time of the giant planet instability (e.g., Parker et al. 2011, Morbidelli et al. 2008), part of which was captured in the Trojan region. Thus the statistical match between the size distributions of Trojans and hot classical KBOs is important observational support of the prediction of the Nice model.

Grav et al. (2011) detected no significant difference in the size distributions of the leading and trailing cloud, beyond the well-established observation that the leading cloud has significantly more objects than the trailing cloud, but they did not sample the small sizes at which Yoshida and Nakamura (2008) noticed the difference between the two swarms. Grav et al. (2011) estimated the fraction of objects with diameter larger than 10 km to be N(leading)/N(trailing) = 1.4±0.2, which is lower than but consistent with previous estimates of 1.6±0.1 derived by Szabó et al. (2007).

As discussed below (section 2.4), Trojans separate spectrally into two groups: a "redder" group with a steep red spectral slope and a "less-red" group with a shallower spectral slope. Using absolute magnitudes (H) from the Minor Planet Center online list of Jupiter Trojans (MPC) and colors from the Sloan Digital Sky Survey (SDSS) for object with H < 12.3, Wong et al. (2014) found that the two spectral groups have distinct magnitude distributions (and therefore likely distinct size distributions, given the relatively uniform albedos among Trojans). Both distributions have a break in slope near H~8.5, just like the total Trojan population. The redder spectral group, however, has a shallower power-law slope on both the bright (large) and faint (small) side of the break than the less-red group, but the difference is greatest on the faint (small) side. Grav et al. (2012) also point out a potential trend in fraction of the two spectral/color groups with size from WISE data, and DeMeo and Carry (2014) report similar changes in abundances of the two spectral groups in their SDSS taxonomy. Wong et al. (2014) suggest that the different power-law slopes indicate that the two spectral groups formed in different regions of the solar nebula, and likely also point to different collisional evolutions before being captured in to Jupiter's Lagrange regions. Alternatively, a scenario in which redder objects are collisionally modified into less-red objects may also be consistent with the data.

### 2.2. Albedos

Over the first few decades of physical studies of the Trojan asteroids, thermal-infrared radiometric observations of a handful of large Trojans (D ≳ 70 km) from ground-based telescopes (Cruikshank 1977, Fernandez et al. 2003) and space-based surveys (IRAS, Tedesco et al. 2002; AKARI, Usui et al. 2011) revealed visible geometric albedos ($p_V$) of only a few percent, making them among the darkest objects in the Solar System. The NEOWISE project (Mainzer et al. 2011) obtained thermal measurements of more than 1700 known Trojan asteroids during its main cryogenic operations from January to October, 2010 (Fig. 2; Grav et al. 2011, 2012). This represented an order of magnitude increase over all previous publications. The NEOWISE observed sample covers almost all of the largest objects, providing a sample that is more than 80% complete down to about 10 km. The albedo distribution derived from NEOWISE over this size range is remarkably stable, having a mean albedo of 0.07±0.03 across all sizes, consistent with C, P, and D taxonomic classes (see section 2.4). This average albedo is somewhat higher than found by previous studies, likely a result of different observing and analysis techniques. Nevertheless, the Jovian Trojan population is one of the darkest populations in the Solar System. NEOWISE detected no difference is evident in the albedo distribution of the leading and trailing clouds. There is also no statistical difference in the albedo distributions of the two spectral groups described below (Grav et al. 2012, Emery et al. 2011).

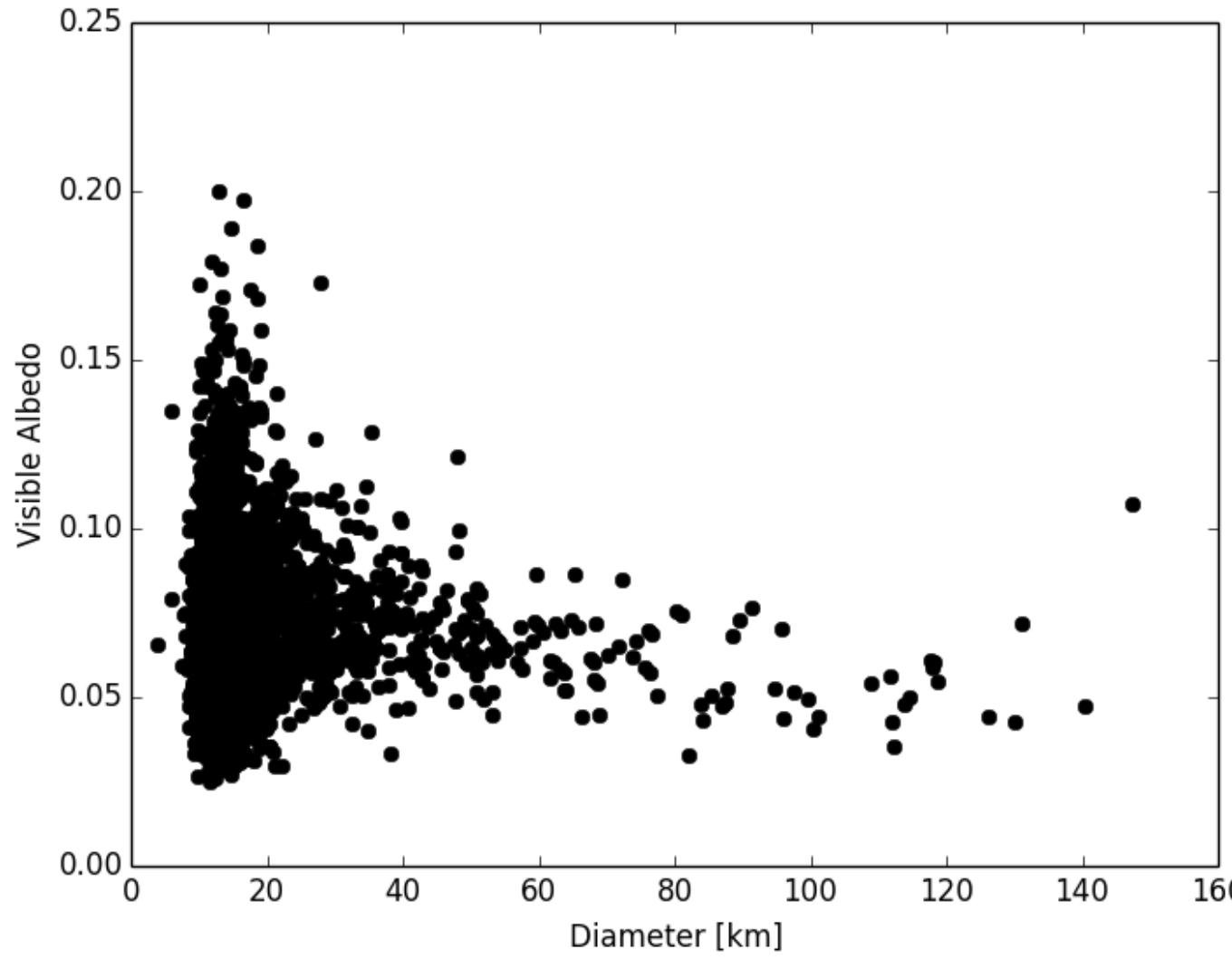

***Figure 2.*** *The diameter versus albedo distribution of the Jovian Trojan population. From Grav et al. (2012).*

Whereas the Spitzer-based observations of Fernandez (2009) suggested an increase in albedo at sizes smaller than 20 km, the NEOWISE observations revealed no such trend. The average albedo of Trojans remains constant down to the smallest objects observed by NEOWISE (~10 km). Higher albedo points in Fig. 2 represent the tail end of what appears to be a gaussian distribution of uncertainties centered on the mean albedo of the entire population (Grav et al. 2011, 2012). It is not expected that any of the small objects really have high albedos.

Note that the NEOWISE-derived albedo of the largest Trojan, 624 Hektor, is significantly higher (0.107±0.011) than derived previously (0.022 to 0.057) using simultaneous visible and infrared photometry (Hartmann and Cruikshank 1978, 1980, Fernandez et al. 2003) and Spitzer Space Telescope infrared spectroscopy (Emery et al. 2006). Hektor is known to be a either an elongated body or contact binary, and the NEOWISE observations showed peak-to-peak amplitude of ~1 magnitude over the 27 hours from first to last observation. Caution should be used in radiometric interpretations of albedo without simultaneous visible photometry, particularly for objects like Hektor that are highly elongated and have large obliquities. If the NEOWISE albedo turns out to be correct, then Hektor would be remarkable not only as a contact binary with moonlets (Marchis et al. 2014), but also as having an anomalously high albedo among the large Trojans.

Fernandez et al. (2003) reported an anomalously high albedo of 0.13 – 0.18 (depending on model parameters) for 4709 Ennomos, which they suggested might be from a recent impact excavating down to a subsurface ice layer. NEOWISE, AKARI, and IRAS all report radiometric albedos of around 0.075 for Ennomos, and Shevchenko et al. (2014) report occultation and phase curve observations from which they derive an albedo of 0.054. Yang and Jewitt (2007) see no evidence for absorptions due to $H_2O$ in near-infrared spectra of Ennomos observed on three different nights. Unfortunately, since the rotation period is very close to 12 hrs (12.2696±0.0005hr; Shevchenko et al. 2014), they would have been observing nearly the same hemisphere each night. It remains an open question whether Ennomos has a bright spot on its surface.

### 2.3. Rotational states and phase curves

Studies of asteroid lightcurves provide information about important properties such as rotation rates, shape, pole orientation, and surface properties. Rotation properties of Main Belt asteroids (MBAs) have been shown to vary dramatically with size (Pravec and Harris 2000; Warner et al. 2009). The rotation of MBAs larger than ~50 km in diameter seems to be determined largely by collisions, while that of smaller bodies is shaped primarily by YORP forces and torques (Pravec et al. 2008). Comprehensive studies have shown that MBAs smaller than ~10 km in diameter are governed by a "spin barrier" corresponding to a rotation period of ~2.2 hours (summarized by Warner et al. 2009). Because of their greater heliocentric distance and low geometric albedos, the Trojans have been less studied until recently.

The orbital eccentricities of the Jovian asteroids are low, with a mean value of 0.074 ± 0.04 (Mottola et al. 2014). They are thus physically isolated from frequent dynamical interactions with other major asteroid groups. While collisions dominate the rotation periods and shapes of large MBAs, factors such as cometary outgassing, tidal braking, and YORP may be significant for the Trojans. Early work by French (1987), Hartmann et al. (1988), Zappala et al. (1989) and Binzel and Sauter (1992) suggested that larger Trojans might have, on average, higher amplitude lightcurves (meaning more elongated shapes) than MBAs of a similar size. All these studies, however, were limited to different degrees either by small sample size or by observational biases favoring large amplitudes and short periods. Because determination of the true shape, surface scattering properties, and pole direction of an asteroid requires observations at many aspect angles, most recent studies have focused on rotation periods rather than systematic coverage of lightcurve amplitudes and determination of pole directions. We focus first on studies of rotation periods, and will conclude with what is known about amplitudes and surface properties.

The past decade has brought the publication of several studies dedicated to eliminating observational bias in Trojan rotation data. Molnar et al. (2007) and Mottola et al. (2011) investigated medium to large Trojans (60-180 km in diameter), while French et al. (2011, 2012, 2013), Stephens et al. (2012, 2014), and Melita et al. (2010) have focused on Trojans less than 60 km in diameter. All investigators have concluded that a significant population of Trojans rotates slowly, with periods greater than 24 hours. Mottola et al. (2014) compared Trojan and Main Belt asteroids in the size range 60 to 180 km; a Kuiper nonparametric statistical test rejects the hypothesis that the two samples belong to the same population at the 5% significance level. For smaller Trojans, the overabundance of slow rotators is even more pronounced. Figure 3, from French et al. (2015), shows the distribution of rotation rates for Trojans less than 30 km in diameter, along with the best fit Maxwellian curve. The Maxwellian is the distribution that would be expected if the spin vectors were oriented isotropically, with each component of the angular velocity following a Gaussian distribution. The curve has been normalized to 1 at the geometric mean rotation frequency for the sample of f = 1.22 revolutions/day (P = 19.7 hours). The excess of slow rotators is obvious.

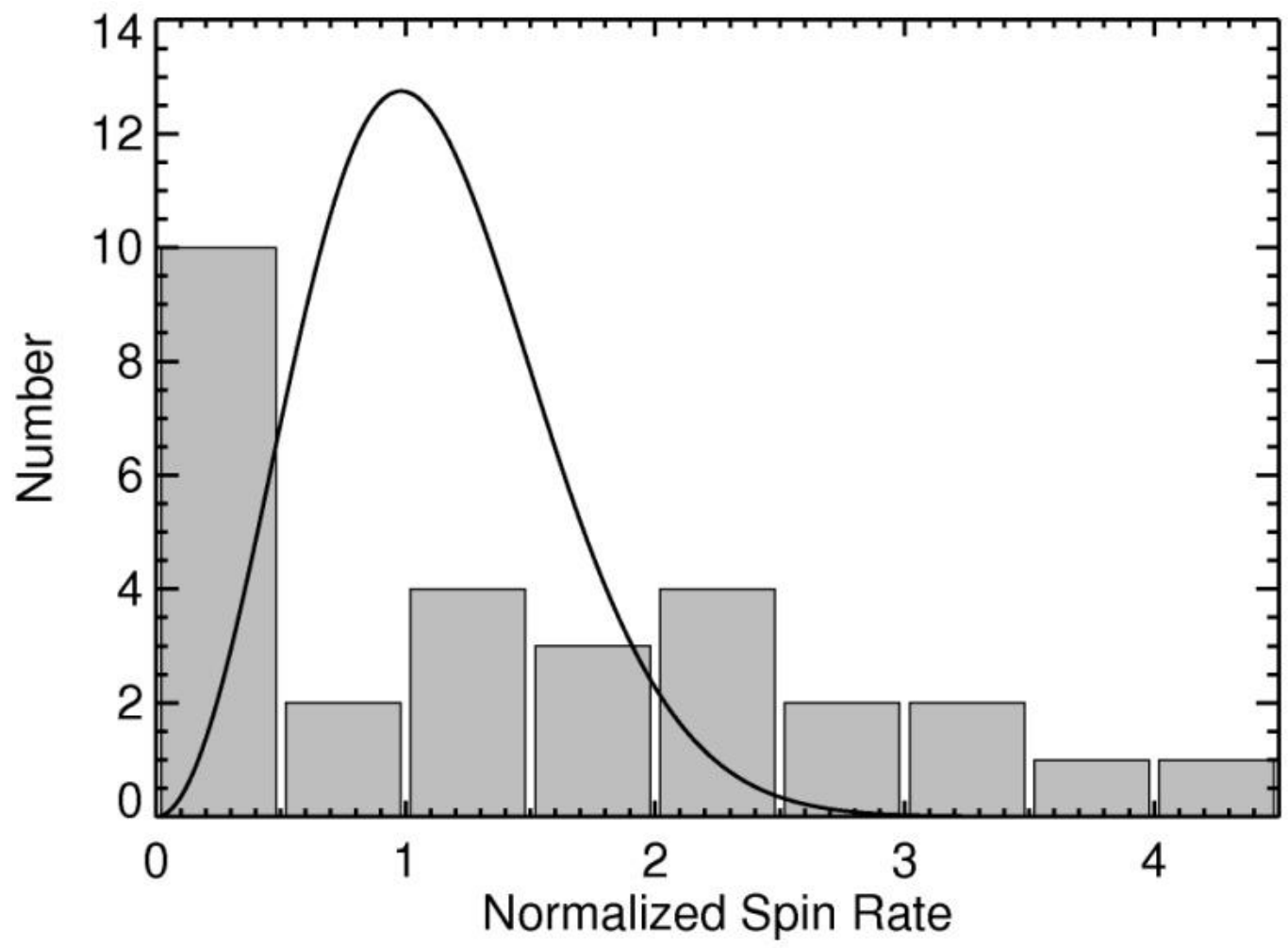

***Figure 3.*** *Distribution of rotation frequencies of 31 Trojan asteroids with D < 30 km versus the best-fit Maxwellian curve. Frequencies have been normalized to the geometric mean for this group of <f> = 1.22 rotations per day (<P> = 19.8 hours). From French et al. (2015).*

The presence of large numbers of slow and fast rotators has already been observed in MBAs, particularly at small diameters. Pravec et al. (2007), in their study of 268 small MBAs, demonstrated that the observed distribution of rotation frequencies is consistent with the Yarkovsky-O'Keefe-Radzievskii-Paddack (YORP) effect as the controlling mechanism. The YORP effect causes a prograde-rotating asteroid to speed up in its rotation and a retrograde rotator's rotation to slow. Because the YORP effect scales as ($R^2/a^2$), where R is the radius of the asteroid and a is the semi-major axis of its orbit, a Trojan asteroid would be affected by YORP to a similar degree as an MBA that is about twice as large. The slow rotation of MBAs as large as 253 Mathilde, at R = 26 km, is thought has been suggested to be caused by YORP (Harris 2004). Thus, Trojans with radii in the 10-15 km range (D = 20-30 km) might be expected to show evidence of YORP, and the large numbers of slow rotators in the leftmost bin of Fig. 3 suggest that they are.

What about fast rotators? The presence of a "spin barrier" at P ~ 2.2 hours has been well documented for MBAs. This represents the critical rotation period, $P_C$, at which a body without internal material strength – a rubble pile – would be spun apart by its centripetal acceleration. This period is

$$P_C \sim 3.3\sqrt{\frac{(1+A)}{\rho}},$$

where $P_C$ is in hours, A is the lightcurve amplitudes in magnitudes, and ρ is the bulk density of the body (Pravec and Harris 2000). Figure 2 of Mottola et al. (2014) shows some evidence for an excess of fast rotators over the MBA population in the 60-180 km range. The French et al. study (in press) includes 31 well-determined lightcurves for sub-30 km Trojans. Currently, no Trojan has been found with a period shorter than that of (129602) 1997WA12 (D = 12.5 km) at 4.84 hours (French et al. 2015). Several other Trojans have periods in the ~5 hour range (Mottola et al. 2014; French et al. 2015). The observed lightcurve amplitudes give density estimates of ~0.5 gm/cm3 if the objects are spinning at the critical period. This value would be consistent with observed comet densities (Lamy et al. 2004). More observations of Trojan rotation periods are encouraged in order to locate the Trojan spin barrier, setting a limit on Trojan densities.

The most recent survey of Trojan asteroid lightcurve amplitudes remains that of Binzel and Sauter (1992). After correcting for the likely bias in published lightcurves due to incomplete sampling at all viewing angles, they concluded that the larger Trojans (D > 90 km) have higher average amplitudes, implying a more elongated shape than MBAs in the same size range. What this means in terms of the evolutionary and collisional history of the Trojans is as yet unexplained.

Most Solar System bodies without atmospheres show an opposition effect (OE) – a sharp, nonlinear brightening near zero phase angle. (The phase angle is the angle between the Sun and the Earth, as seen from the object. For the Earth's Moon this corresponds to Full Moon). High-quality asteroid phase curves generally show a linear slope between phase angles of 5 and 25°, with differing slopes for different albedo asteroids (Belskaya and Shevchenko 2000). At phase angles less than 5°, an opposition surge is observed; this is now understood as due to coherent backscattering, as it is stronger for higher albedo surfaces (Muinonen *et al.* 2002). Phase curves for Trojan asteroids are linear down to phase angles of ~0.1-0.2° (Shevchenko et al. 2012). This linear behavior differs dramatically from the sharp opposition spikes seen in several Centaurs, and is similar to what is observed for dark outer Main Belt and Hilda asteroids (Shevchenko et al. 2012). Shevchenko et al. (2012) attribute the absence of a strong opposition surge to the low albedos of Trojan asteroids. For such low albedos, multiply scattered light, which is required for the coherent-backscatter opposition effect to occur, does not provide a significant contribution to the reflected flux.

### 2.4. Spectral Properties

The first visible-wavelength reflectance spectra of Trojan asteroids were featureless, but the relatively steep, red spectral slopes were excitingly interpreted to indicate the presence of abundant complex organic molecules on the surfaces, masking an ice-rich interior (Gradie and Veverka 1980). Over the following two decades, reflectance spectroscopy at visible and near-infrared (VNIR; 0.4 to 4.0 μm) wavelengths continued to show a range of spectral slopes, but no absorption features (see Dotto et al. 2008), placing strong constraints on the presence of ice near the surfaces and on the presence and form of organic material. Recent dedicated spectral searches for ices in the Eurybates family (DeLuise et al. 2010), on several large Trojans, including Ennomos (Yang and Jewitt 2007), and on several of the smaller (D ~ 10 to 30 km) Trojans for which the NEOWISE survey suggests high albedos (Marsset et al. 2014), as well as a general NIR survey (0.7 to 2.5 μm; Emery et al. 2011) still reveal no spectral absorption bands. Yang and Jewitt (2011) re-observed 7 large Trojans whose spectra had hinted at a possible broad 1-μm silicate band, but those too turned out to be featureless.

Statistical analyses of VNIR colors and spectra have revealed the presence of two distinct spectral groups (Fig. 4), a "red" group consistent with the asteroidal D-type taxonomic class and a "less-red" group consistent with the asteroidal P-type classification (Szabó et al. 2007, Roig et al. 2008, Emery et al. 2011**,** Grav et al. 2012). Emery et al. (2013) supplemented the NIR sample with 20 additional L5 Trojans, showing that the two spectral groups appear to be equally distributed in the two swarms. The NIR sample is restricted to objects larger than ~70 km, and it is not yet clear if the bimodality extends to smaller sizes (e.g., Karlsson et al. 2009). Emery et al. (2011) suggest that the spectral groups represent two compositional classes that potentially formed in different regions of the solar nebula. Otherwise, no strong correlations between spectral and any physical or orbital parameter are present (Fornasier et al. 2007, Melita et al. 2008, Emery et al. 2011), though Szabó et al. (2007) suggest a weak correlation with orbital inclination in the L4 swarm that Fornasier et al. (2007) attribute to the presence of the Eurybates family. Brown et al. (2014) presented spectra in the 2.85-4.0 μm region showing a possible absorption for a few "less-red" Trojans similar to that seen on 24 Themis (Campins et al. 2010, Rivkin and Emery 2010). The objects that Brown et al. observed from the "red" group showed no absorption.

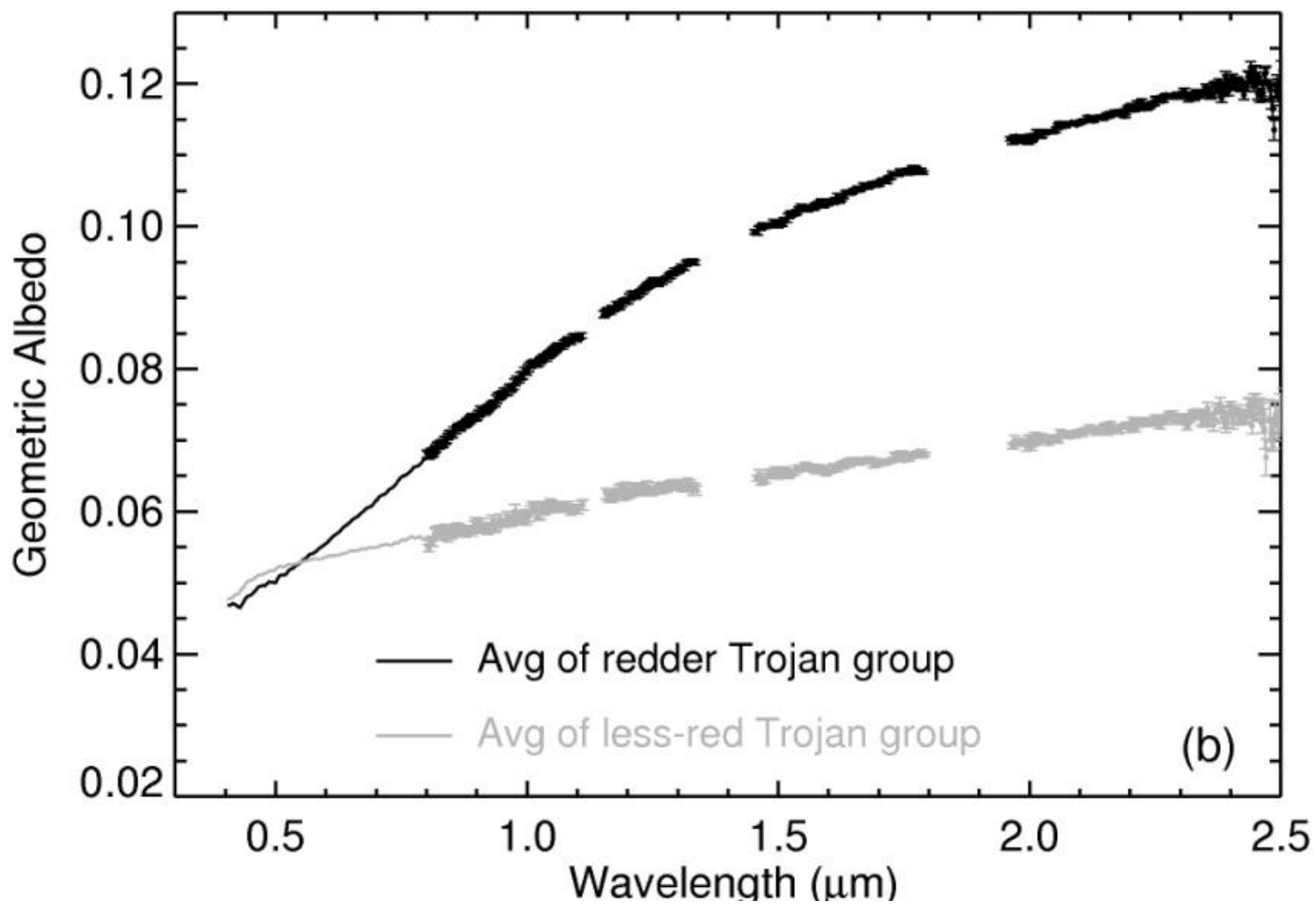

***Figure 4.*** *Combined visible and NIR average spectra of the two spectral groups. The spectral groups are separated more clearly when both visible and NIR wavelength ranges are considered. These spectra have been scaled to $p_v$=0.055. From Emery et al. (2011).*

Mid-infrared (MIR; 5 – 38 μm) emissivity spectra have been published of four Trojan asteroids (Hektor, Agamemnon, Aneas, and Patroclus), and all four show strong emissivity peaks near 10 and 20 μm (Emery et al. 2006, Mueller et al. 2010). It is interesting to note that although the emissivity features seen in Patroclus, the only "less-red" object among the four, are in the same location as for the other three Trojans, the spectral contrast is significantly weaker. Whether this is a trend that follows the spectral groups remains to be discovered. From mutual eclipses of the binary components, Mueller et al. (2010) derived a very low thermal inertia (~6 – 20 J $m^{-2}K^{-1}s^{-1/2}$) for Patroclus. Thermal spectral energy distributions of other (large) Trojans are also consistent with very low thermal inertia surfaces (e.g., Fernandez et al. 2003, Emery et al. 2006), suggesting very fine-grained, porous regoliths. Horner et al. (2012) computed a slightly higher thermal inertia of 25 to 100 $m^{-2}K^{-1}s^{-1/2}$ for Anchises, but still consistent with a "fluffy" regolith.

### 2.5. Binarity / Densities

Binaries provide invaluable data about the physical nature of asteroids. Two are presently known amongst the Trojans, and they present intriguing comparisons. 617 Patroclus has a less-red surface, and the two components are nearly equal in size (Merline et al. 2002). The bulk density of the components is 1.08 ± 0.33 g/$cm^3$ (Marchis et al. 2006a). The orbit is nearly circular, and the rotation periods appear to be synchronized with the orbital motion, implying that the bodies are in a principal-axis rotation state (Mueller et al. 2010). The 102.5 hr period is well explained by tidal braking. 624 Hektor, on the other hand, has a rotation period of 6.924 hours and appears to be either a contact binary or one extremely elongated object with a small moon of diameter ~12 km (Marchis *et al.* 2014). Its bulk density has been determined to be 1.0 ±0.3 g/$cm^3$ (Marchis et al. 2014), very close to that of the Patroclus system. Hektor has a redder, spectrum (Emery et al. 2011), suggesting a possible difference in composition. Analysis of the Hektor system suggests a high-inclination (~166°) and high-eccentricity (~0.3) orbit for the satellite, with an orbital period just between two spin-orbit resonances. This implies that the orbit has not evolved significantly since the formation of the system and is therefore primordial (Marchis *et al.* 2014). Most recently, Decamps (2015) reanalyzed lightcurve data and adaptive optics images of the Hektor contact binary in terms of a dumbbell shape, finding a better fit to the data and a smaller volume than the previous shape model. This smaller volume results in a higher density estimate of 2.43±0.35 g cm$^{-3}$. Hektor and Patroclus may therefore have different internal structures as well as belonging to different spectral groups.

Searches for other Trojan binaries have been undertaken by several researches. In a study of lightcurve amplitudes, Mann et al. (2007) report two objects with lightcurve amplitudes of ~1 mag (17365 1978 VF11 and 29314 Eurydamas) and suggest these might be contact binaries. From their survey of 114 Trojans, they estimate that 6–10% of Trojans might be contact binaries. While observing a stellar occultation by Agamemnon, Timerson et al. (2013) detected a brief dip after the main occultation, which they interpret as a potential moonlet. Most recently Noll *et al.* (2013) observed eight Outer Main Belt and Trojan asteroids with long rotational periods. No binaries were found, and those authors concluded that binaries are less frequent in the Outer Main Belt and Trojan regions than in the Kuiper Belt.

### 2.6. Physical Interpretation of Observations

In some ways, it seems that the Trojans are conspiring to keep the secret of their compositions and physical structure hidden. Nevertheless, the persistent effort of characterization described in the previous sections is paying off. The clearest indication of internal structure comes from the determination that Patroclus and Hektor both have bulk densities near 1 g cm$^{-3}$. This low density, relative to rock and even carbonaceous chondrites, indicates either a significant low density component (i.e., ice), a large macroporosity, or, more likely, a combination of the two. The distribution of rotation rates and sizes have both been used to argue for a division in which the largest Trojans (D > 80 to 130 km) are intact, primordial planetesimals, whereas the smaller bodies are collisional fragments (Binzel and Sauter 1992, Jewitt et al. 2000, Yoshida and Nakamura 2005, 2008, Grav et al. 2011, Fraser et al. 2014). If the internal compositions are distinct from surface compositions (i.e., if a surface crust hides an ice-rich interior), one would expect the properties of smaller Trojans to be systematically different from those of larger Trojans. The small Trojans are at the limit of current observing capabilities from most characterization techniques, but there does not appear to be a systematic difference between large and small Trojans.

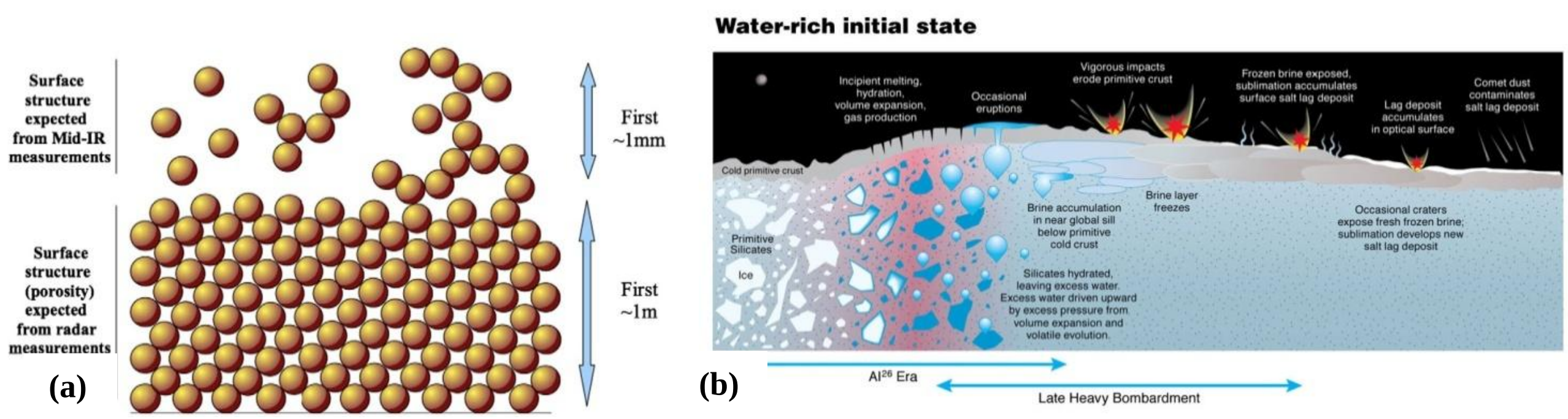

***Figure 5.*** *(a) Schematic model of an underdense, "fairy-castle" regolith on Trojan asteroids deduced from comparisons between MIR emissivity spectra of Trojans and laboratory measurements of powdered meteorites mixed with KBr. (From Vernazza et al. 2012). (b) Evolutionary scenario that might produce salt-rich surfaces, in which embedded fine-grained silicate dust could explain measured MIR emissivity spectra of Trojans (From Yang et al. 2013).*

The featureless VNIR spectra can be used to assess what is *not* on Trojan surfaces, but do not give a clear indication of what *is* on these surfaces. The red VNIR slopes have often been cited as suggestive of abundant organic material. However, Emery and Brown (2004) argue that the absence of strong absorptions in the 2.85 – 4.0 μm spectral range strongly limits the types and abundance of organics, and that, therefore, the spectral slopes cannot be due to organics. Rather, they and Emery et al. (2011) demonstrate that the featureless, low-albedo, red-sloped VNIR spectra can be fit by amorphous and/or space weathered silicates. Spectral models have been used to place upper limits of only a few wt% of $H_2O$ ice on the surfaces (e.g., Emery and Brown 2004, Yang and Jewitt 2007).

The MIR emissivity spectra that have been published demonstrate convincingly that Trojan surfaces are populated by silicate dust. The large spectral contrast and positive polarity (i.e., that the features appear as peaks rather than valleys) indicates that the dust is very fine-grained ($\lesssim$ 10 μm-sized grains) and that the grains are fairly well separated (Emery et al. 2006). No cometary (extended) emission has been detected around Trojans, so these spectra provide constraints on the surface structure. Vernazza et al. (2012) investigated a model in which the regolith is very porous (i.e., an extreme "fairy-castle" structure; Fig 5a) using laboratory measurements of meteorites powders mixed with KBr. Their experiments demonstrate the viability of reproducing the MIR spectra, and they find that the features indicate dust composed primarily of amorphous forsteritic olivine, but with a non-negligible crystalline fraction as well. This model is consistent with the very low thermal inertias measured for Trojans. Yang et al. (2013), on the other hand, envision a surface where silicates are embedded in a transparent matrix. They demonstrate, with laboratory measurements and spectral modeling, that salts could provide the matrix, and discuss possible evolutionary scenarios (Fig 5b). In either case, the MIR emissivity spectra point to a silicate fraction that is compositionally similar to cometary silicates (Emery et al. 2006, Vernazza et al. 2012).

## 3. Origin and Evolution

### 3.1. Origin of Jupiter Trojans

The capture mechanisms proposed so far for explaining the presence of Trojan populations in the Lagrange regions of the planets can be broadly divided into two main classes:

1) Trapping due to non-gravitational perturbations on primordial planetesimals passing by the planet. Trapping can occur because of
    a) Drift into the Trojan region due to the action of a dissipative force like gas drag (Yoder 1979, Peale 1993) or the Yarkovsky effect. These processes affect small bodies, which could have subsequently grown into larger asteroids once trapped in tadpole orbits.
    b) Collisions occurring close to the resonance border which can inject fragments into Trojans orbits (Shoemaker et al. 1989).
2) Changes in the physical and orbital parameters of the planet can lead to a shift in the position of the Lagrangian points causing the capture of local planetesimals. Four specific mechanisms have been proposed.

a) Mass growth of the planet. This causes an expansion of the resonant area, capturing close-by planetesimals as Trojans (Marzari and Scholl 1998a,b; Fleming & Hamilton 2000).
b) Smooth migration of the planet. Objects are swept into the Lagrange regions (Lykawka et al. 2009)
c) Crossing of a mean motion resonance of the planet with another planet. A chaotic path can be opened during the evolution due to secular resonance sweeping and the superposition of secondary resonances between harmonics of the Trojan libration frequency and the critical argument of mean motion planetary resonance. Planeteseimals can be trapped in tadpole orbits via this chaotic path which is closed once the resonance crossing (Morbidelli et al. 2005, Marzari et al. 2007). The crossing of the 2:1 resonance between Jupiter and Saturn has been invoked in the Nice model to explain the capture of the Jupiter Trojans.
d) Jumping Jupiter. A period of instability of the planet orbit due to close encounters with a second planet causes steps in semi-major axis which may lead to the capture of leftover planetesimals due to their sudden dislocation within the stable tadpole regions of the planet (Nesvorný et al. 2013). This might explain an asymmetry between L4 and L5 since the perturbing planet may temporarily cross the Trojan region dispersing a consistent fraction of the local population.

The early models on the origin of Jupiter Trojans are reviewed in the chapter by Marzari et al. (2002). In essence, they considered capture by gas drag or the pull-down process, which is due to the broadening of the tadpole region around the Lagrange triangular points occurring during the increase of the mass of the planet. Both these models have several problems in reproducing the observations, the most severe of which is the inclination distribution. Jupiter’s Trojans cover the inclination range 0-35 degrees, with a median inclination of 10 degrees (which becomes 18 degrees for bright Trojans with $H<12$, for which our knowledge of the population is bias free), while the aforementioned capture models had problems explaining any significant inclination excitation.

It is worth noting that for the Trojan population the eccentricity excitation is much less than twice the inclination excitation (the relationship expected for a randomly excited disk). In fact, the eccentricities are smaller than 0.15, with a few exceptions. But this is due to the boundaries of the stability region. Levison et al., (1997) mapped these boundaries with long-term numerical simulations and demonstrated that the Trojans fill the entire region that is stable over the age of the solar system.

In 2005, Morbidelli et al. proposed a radically different model for the origin of the Trojans, developed in the framework of a scenario later named “Nice model”. In the original version of the Nice model, the giant planets were originally in a more compact configuration on quasi-circular and co-planar orbits. The planets migrated slowly in divergent directions as they scattered planetesimals, originally located beyond Neptune’s orbits. As the initial ratio of the orbital periods of Saturn and Jupiter was postulated to be slightly less than 2, the divergent migration brought these planets to cross their mutual 1:2 mean motion resonance. This resonance crossing excited the eccentricities of Jupiter and Saturn and destabilized the planetary system as a whole. A phase of close encounters among the planets followed, with Uranus and Neptune scattered outwards onto large eccentricity orbits. Thus Uranus and Neptune dispersed the original trans-Neptunian disk and, by a feedback process, all planetary eccentricities were damped to moderate values, consistent with the current ones and the giant planet system eventually the current orbital configuration. In this model, the capture of Trojans occurred during the 1:2 resonance crossing. In fact, the tadpole region becomes fully unstable when the planets are near this resonance. This means that the planetesimals scattered from the trans-Neptunian region can enter and exit the tadpole region. But when Jupiter and Saturn migrate far enough from the 1:2 resonance, the tadpole region becomes suddenly stable. The planetesimals that are there at that time are then trapped forever. A detailed map of the stability of the tadpole region as a function of the Saturn/Jupiter period ratio can be found in Robutel and Bodossian (2009).

Thus, the Morbidelli et al. (2005) paper was the first prediction of capture of Jupiter Trojans from the trans-Neptunain disk. The simulations allowed reproducing, at least qualitatively, the distribution of the observed Trojans in eccentricity, inclination and libration amplitude. The capture probability into the Trojan region was shown to be large enough to justify the currently observed population, starting from a primordial trans-Neptunian disk of 50 Earth masses, with a Kuiper-belt like size-frequency distribution.

The original version of the Nice model, however, proved to be not entirely satisfactory. Further investigation of the dynamics of the giant planets in the primordial disk of gas showed that the giant planets should have emerged from the gas-disk phase locked in mean motion resonances with each other (Morbidelli et al., 2007; Walsh et al., 2011; see Morbidelli 2013 for a review). The instability of the planetary system than occurred when two planets fell off resonance, under the perturbations of the planetesimal disk and not when Jupiter and Saturn crossed their 1:2 mean motion resonance (Levison et al., 2011). Also, of all the possible evolutions that the giant planets can follow during the instability phase, it was shown that the only acceptable ones are of “jumping-Jupiter type”, which are evolutions where Jupiter scatters outward a planet (Uranus, or Neptune or a rogue fifth planet of comparable mass) previously scattered inwards by Saturn. In this case, the period ratio between Saturn and Jupiter jumps up, impulsively. This is needed because otherwise the slow increase in the orbital separation between Saturn and Jupiter drives secular resonances across the asteroid belt (Morbidelli et al., 2010) and the terrestrial planets region (Brasser et al., 2009), leaving both populations on orbits inconsistent with the current ones. For this not to happen, the period ratio between Saturn and Jupiter has to jump from the original value of ~1.5 (the 2:3 mean motion resonance) to more than 2.3. This has become a basic requirement of success for the modern Nice model simulations (see for instance Nesvorný and Morbidelli, 2012). In the case, however, there is no 1:2 resonance crossing and the original Trojan capture model of Morbidelli et al. (2005) is invalidated.

Nesvorný et al. (2013) have re-investigated the possibility of capture of Jupiter Trojans in the framework of the jumping-Jupiter scenario. They did this using three simulations of evolution of the giant planets (all starting initially with 5 planets, see Nesvorný and Morbidelli, 2012) satisfying all constraints, particularly the jump of the period ratio to a value larger than 2.3, with a residual migration not driving the period ratio beyond 2.5 (the 2/5 resonance). These planetary evolutions are shown in Fig. 6

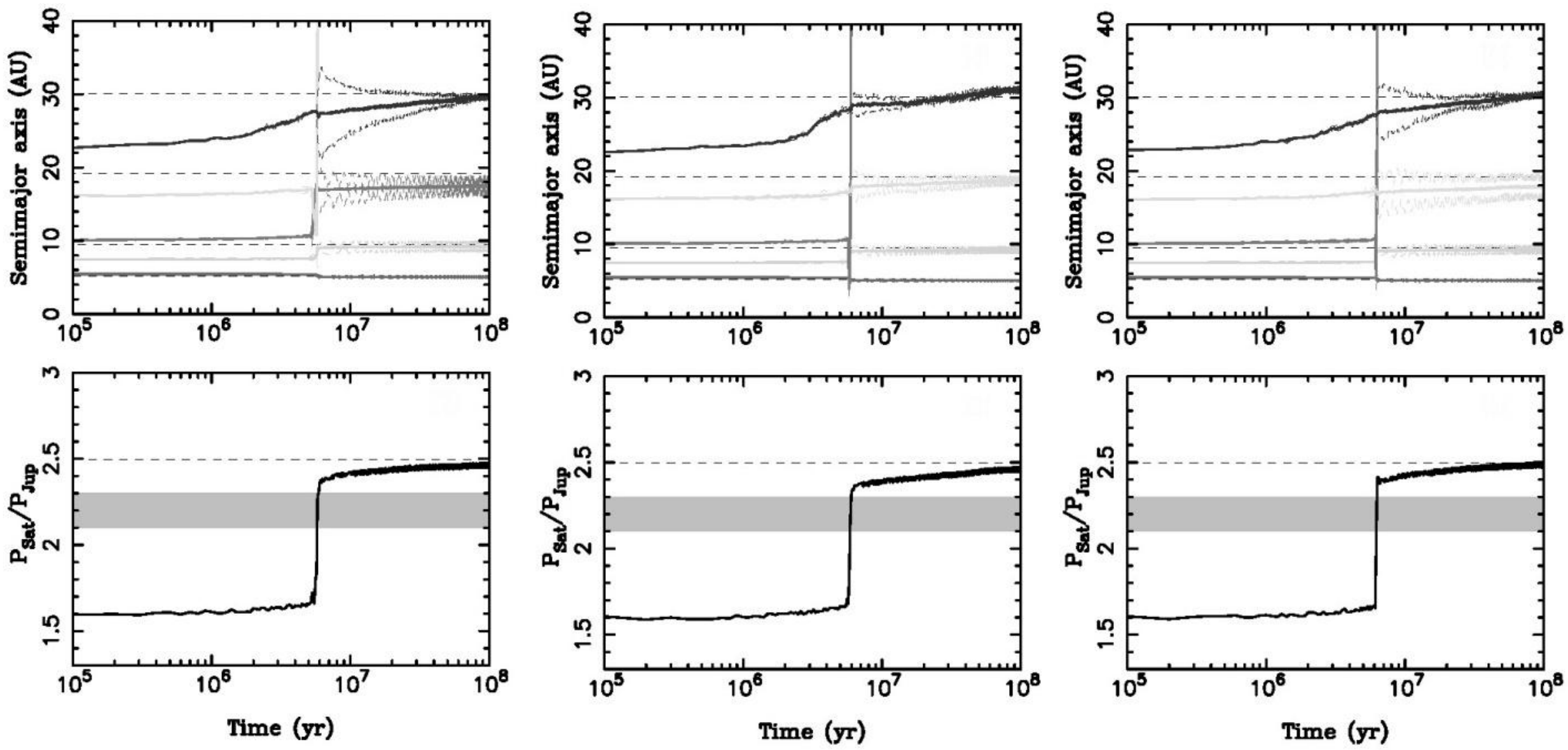

*Fig. 6. The evolution of the giant planets in the three simulations considered in Nesvorný et al. (2013) for the capture of Jupiter's Trojans. The top panels show the evolution of semi major axis (solid), perihelion and aphelion distances (dashed) for Jupiter (bottom, dark gray), Saturn (2$^{nd}$ from bottom, light gray), Uranus (middle, dark or light gray, depending which remains) , Neptune (top, dark gray) and the planet ultimately ejected from the system (middle, light or dark gray, depending which is ejected). The lower panels show the evolution of the period ratio between Jupiter and Saturn. The gray band shows the forbidden region, corresponding to secular resonances in the asteroid belt or in the terrestrial planet region.*

They found that Trojans can be captured during Jupiter's jump. In essence, the captured planetesimals are those that, by chance, are on a moderate eccentricity orbit just inward of Jupiter's location at the time of the Jump. When Jupiter jumps inward, these planetesimals can then fortuitously find themselves in the tadpole region. This mechanism is illustrated in Fig. 7.

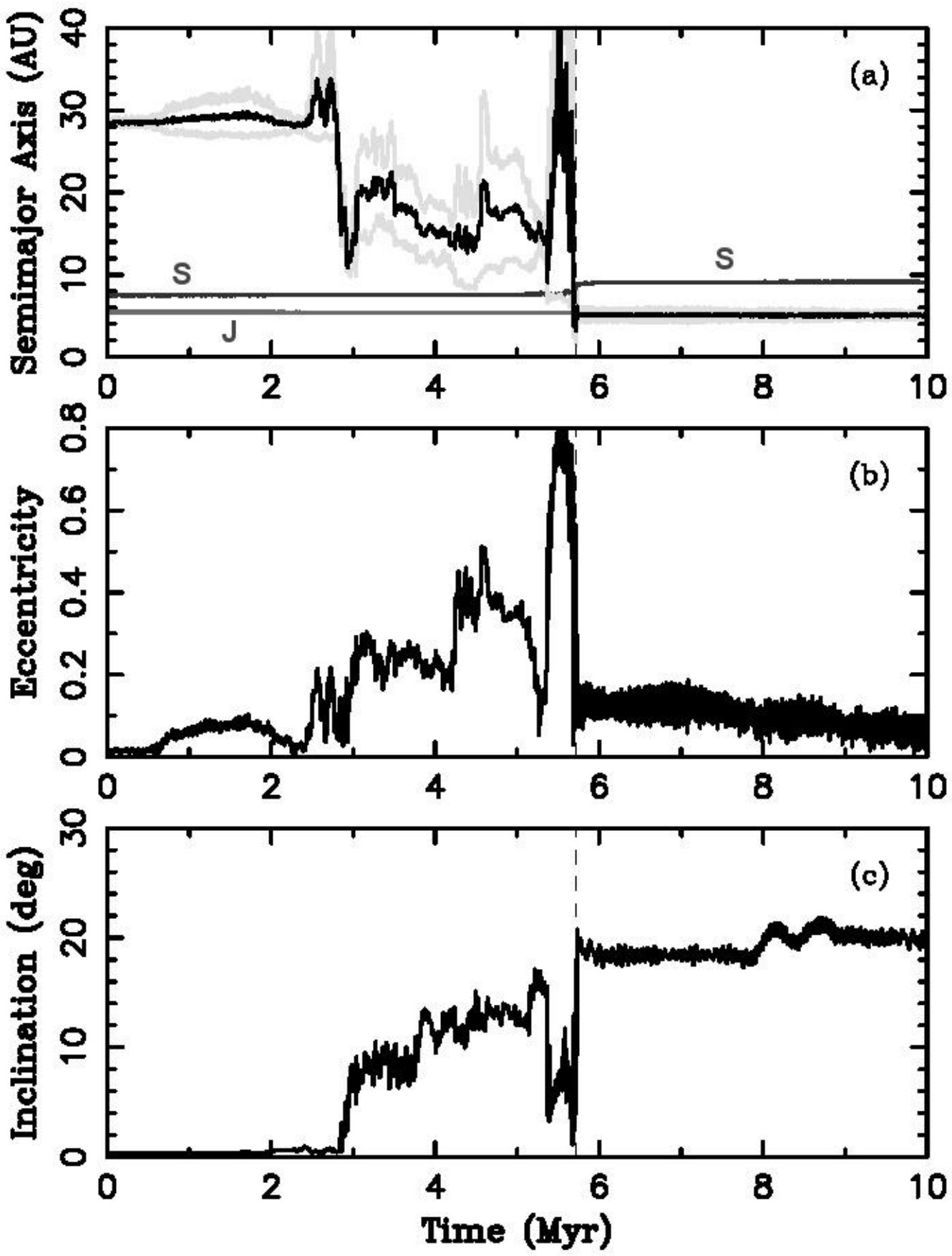

*Fig. 7. An example of capture of a Trojan body. The top panel shows the evolution of semi-major axis of the body in black and of its perihelion and aphelion distances (light gray), together with the semi-major axes of Jupiter (bottom gray curve marked "J") and Saturn (dark gray curve marked "S"). Notice the perfect overlapping of the particle's semi-major axis, perhelion, and aphelion and Jupiter's semi-major axis at the end, proving the capture of a low-eccentricity Trojan. The middle panel shows the evolution of the eccentricity and the bottom panel of the inclination of the captured particle. From Nesvorný et al. (2013).*

Most of the captured planetesimals turned out to be only temporarily unstable, so Nesvorný et al. continued the simulations of the captured bodies over 4 Gy and finally they analyzed the orbits of the Trojans surviving in the tadpole region till the end.

The resulting orbital distribution turned out to be remarkably similar to the observed one. This is illustrated in Fig. 8 through cumulative distributions. Only the inclination distribution seems not very correct, the observed one being less excited than the synthetic one. But the Trojans' distribution is biased towards low inclinations. To remove the bias, Nesvorný et al. considered also the Trojans with H<12, which constitute a complete set (Szabó et al. 2007). The match becomes excellent. This model reproduces the observed distribution even better than the original 2005 model.

In addition to the orbital distribution, there are two qualitative advantages of the new model over the previous one. First, this model has the potential to explain the ~30% asymmetry between the L4 and L5 populations. In fact, unlike the previous model which was strictly symmetric for the two tadpole regions, the new model can capture more or fewer bodies in one of the two clouds depending on the specific geometry of the planetary encounter that causes the jump in Jupiter's orbit. Imagine for instance that the rogue planet passes through one of the two tadpole region coming out of its last encounter with Jupiter and it is intuitive to understand that fewer bodies will remain stable there. Indeed, the three simulations presented in Nesvorný et al., produced asymmetries at the 30-80% level (not necessarily in favor of L4).

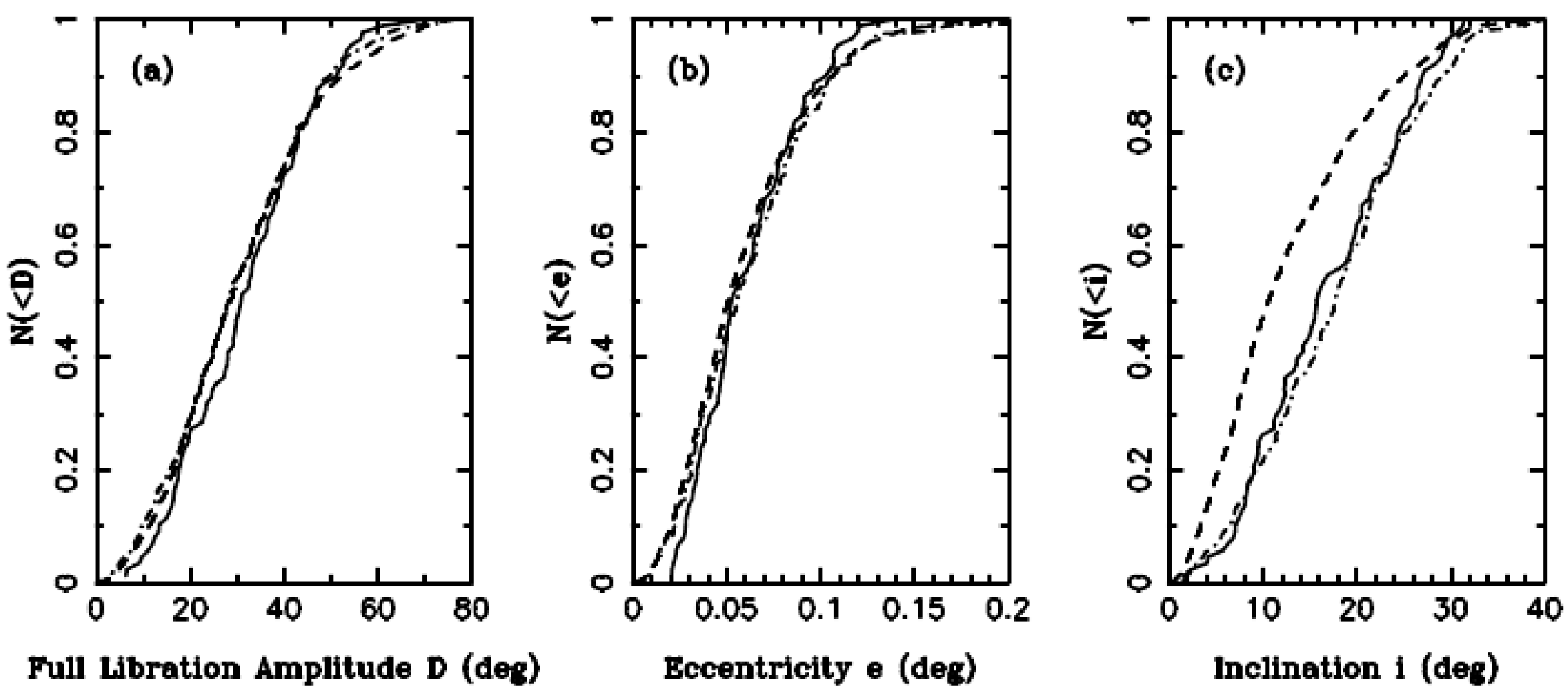

*Fig. 8. The cumulative distribution of real (dashed) and captured Trojans (solid) in libration amplitude (left), eccentricity (middle) and inclination (right). In the right panel, the dash-dotted curve shows the observed inclination distribution of Trojans with H<12. Unlike the dashed-distribution, the latter should be bias free. From Nesvorný et al. (2013).*

The second advantage is that in the new model the capture of Trojans occurs with some time-lag relative to the onset of the instability of the giant planets. The latter occurs when two planets fall out of resonance, while Trojans' capture occurs during the last encounter of Jupiter with the scattered planet, which can occur only once the unstable phase is occurring. Instead, in the original model the capture occurred at the 1:2 resonance crossing which triggered the instability. The difference may be important for the capture of asteroids. If the capture of Trojans occurs at the beginning of the planet instability, the asteroid belt does not have the time to be partially destabilized and therefore asteroids cannot be captured in the Trojan region. This however can occur in the new model. Thus, this opens the possibility that some Trojans, for instance those consistent with the asteroidal C-types, which are a minority of the Trojan population, may come from the outer asteroid belt.

Nesvorný et al. (2013) found that the probability of an original trans-Neptunian object to be captured over the age of the Solar System as a Trojan is 6-8 x $10^{-7}$. This implies 3-4x$10^{7}$ planetesimals with H<9 (D>80km for a=7% albedo) in the original trans-Neptunian disk. With a Kuiper belt-like size frequency distribution, this is consistent with the mass needed in the new version of the Nice model (Nesvorný and Morbidelli, 2013). This disk population is also consistent with the crater record on Iapetus (Rivera-Valentin et al., 2014) showing the overall consistency of the new Nice/jumping-Jupiter model.

Thus, the Nesvorný et al. (2013) model reinforces the idea that Trojans are, in vast majority, objects captured from the trans-Neptunian disk. The same disk gave origin also to the hot Kuiper belt population and the scattered disk (Levison et al., 2008), whereas the cold Kuiper belt population might have been a separate population not significantly affected during the giant planet instability (Parker and Kavelaars, 2012; Batygin et al., 2011; Fraser et al., 2014). Because both the hot population and the Trojan population should not have suffered any significant collisional evolution at large sizes since the giant planet instability time (Levison et al., 2009), Morbidelli et al. (2009) predicted that size distributions of the Trojans and of the hot population should have been the same. At the time, this was a real prediction, because the size distribution of the hot population was known only for sizes much larger than the largest Trojan, and it looked much shallower. So, the prediction was that, at sizes comparable to those of the Trojans (less than 200km in diameter) the size distribution of the hot population would steepen up and look like that of the Trojans. This prediction has been recently supported by Fraser et al. (2014) and confirmed by Adams et al. (2014). We refer the reader to section 2.1 for a more detailed discussion of the Trojans' size distribution.

### 3.2. Stability Properties

The dynamical stability of Trojans is affected by different types of resonances which influence their survival in the present Solar System and may have even played a significant role during the potential migration of planets in early Solar System evolution. These resonances involve the fundamental frequencies of the Trojan motion, which can be related by a ratio of two small integers to those of the planetary system. Approximate analytical expressions have been derived for the frequencies of the Trojan motion within the simplified Elliptical Restricted Three Body Problem (ERTBP). The libration motion around either $L_4$ or $L_5$ is characterized by a long period frequency given by $\nu_l = \frac{1}{2}(1 - \sqrt{1 - 27\mu(1-\mu)})^{\frac{1}{2}} n_p \simeq \sqrt{27/4\,\mu}\, n_p$ and a short period frequency $\nu_s = \frac{1}{2}(1 + \sqrt{1 - 27\mu(1-\mu)})^{\frac{1}{2}} n_p \simeq \sqrt{(1 - 27/4)\mu}\ n_p$ (Erdi et al. 2007, 2009), where $\mu = \frac{m_p}{m_s + m_p}$ is the mass ratio and $n_p$ is the planet mean motion. As an example, for Jupiter's Trojans $T_l \simeq 147.8\, yr$ and $T_s \simeq 11.9\, yr$. The secular frequency of the perihelion longitude precession $g^{ERTBP}$ is analytically given, at the second order in the libration amplitude $d$, as $g^{ERTBP} \simeq (27/8 + (129/2^6)d^2)\mu n_p$(Erdi 1988), while the precession frequency of the nodes is computed as $s^{ERTBP} \simeq 3/4\, d^2 \mu n_p$. In the more general problem of the Trojan motion in the full planetary system, the values of the frequencies $\nu_{l,s}, g, s$ depend on the orbital elements of the Trojan orbit and of the planets. In particular, the secular frequencies $g, s$ will include the contribution of the planets, becoming $g = g^{ERTBP} + \sum_{j \neq p} g_j$ and $s = s^{ERTBP} + \sum_{j \neq p} s_j$, where $g_j$ and $s_j$ are the eigenfrequencies of the classical Lagrange-Laplace solution of the secular problem. Precise semi-empirical expressions have been derived for $g$ and $s$ as a function of the Trojan orbital parameters, fitting the outcomes of direct numerical integrations of Trojan trajectories and of all the planets of the Solar System (Marzari et al. 2002, 2003a, 2003b, 2005). An integer, or near integer, relation between a frequency of the Trojan motion and one or more frequencies of the planets leads to a resonant interaction that can destabilize the tadpole motion. The possible different types of resonances have been grouped (Robutel and Gabern 2006, Erdi et al. 2007, Robutel and Bodossian 2009) into 4 families:

*Family I*: commensurabilities between the orbital frequency of the planet $n_p$and the libration frequencies of the Trojan motion $\nu_l, \nu_s$enriched by additional secular frequencies of the planetary system. They are defined by the expression $i\nu_{l,s} + jn_p = -(kg + ls + \sum_n k_n g_n + \sum_n l_n s_n)$ where $i, j, k, l, k_n, l_n$are integers satisfying the relation $j + k + l + \sum_n k_n + \sum_n l_n = 0$ imposed by the d'Alembert rules.

*Family II*: commensurabilities between $\nu_l, \nu_s$ and the libration frequency $\psi_{p,q}$of the critical angle $\theta = p\lambda_m - q\lambda_n + \ldots$ of a *p:q* mean motion resonance between two planets *n,m* defined as $\psi_{p,q} = pn_n - qn_m$ . In this case the relation among the frequencies becomes $i\nu_{l,s} - j\psi_{p,q} = -(kg + ls + \sum_n k_n g_n + \sum_n l_n s_n)$, with $j(q-p) + k + l + \sum_n k_n + \sum_n l_n = 0$. In the Solar System, an important almost resonance between Jupiter and Saturn is the so called Great Inequality 5:2.

*Family III*: secular resonances between $g, s$and the eigenfrequencies of the Solar System, defined by the condition $kg + ls = -(\sum_n k_n g_n + \sum_n l_n s_n)$ and $k + l + \sum_n k_n + \sum_n l_n = 0$

*Family IV*: commensurabilities between the libration frequency of the planetary resonance p:q and the secular frequency of the Trojan motion $g$defined as $ig + k\psi_{p,q} = -(ls + \sum_n k_n g_n + \sum_n l_n s_n)$ with $j(q-p) + k + l + \sum_n k_n + \sum_n l_n = 0$

Present Jupiter Trojans are perturbed by all these families of resonances. Overlap of these resonances generates extended chaotic regions, which limits the extent of the phase space populated by stable orbits. In Fig. 9, a diffusion map (Robutel and Gabern 2006) shows the stability properties of fictitious Trojan orbits of Jupiter as a function of their initial semi-major axis and eccentricity. The color coding measures the diffusion rate in the phase space computed by means of the Frequency Map Analysis (Laskar 1990), a powerful numerical tool for the detection of chaos from numerical integration. The color scale ranges from blue, corresponding to stable regions, to red for highly chaotic orbits, while in black are displayed those test bodies that are ejected on a short timescale. The red arch limiting the stable region from above is due to the *family III* nodal secular resonance $s - s_6$, as clearly shown in the power spectrum of a Trojan orbit lying close to the arch (Fig. 10; Marzari et al. 2003a). For higher inclinations of the test Trojan orbits, additional secular resonances such as $s - 2s_6 + s_7, 3s - 4s_6 + s_7, 2s - 3g_5 + g_6, 3s - s_6 - 2g_5$ and others come into play, reducing the size of the stable region. Superposition of *family I* resonances is responsible for the large chaotic zone extending beyond 5.35 AU limiting the libration amplitude of Trojan orbits. The $i = 13$ and $i = 14$ *family I* resonances generate the two main v-shaped unstable yellow structures within 5.25 AU. *Family II* resonances, whose influence was also argued by Nesvorný and Dones (2002), are responsible for the finger-like structures extending from the outer layer of the stable region towards small eccentricities in between 5.25 and 5.35 AU. Finally, the thin yellow structures in the small libration region for $a$ ⍰ $5.27$are due to f*amily IV* resonances. When the initial inclination of the orbits is varied, all resonances change location since the main frequencies of the Trojan motion, namely $\nu, g, s$, depend on inclination, but they are still responsible for the main features of the stable regions.

The resonant structure described above evolves during planet migration and can explain the chaotic capture of primordial Trojans in the original version of the Nice model (Morbidelli et al. 2005). In a simplified four body model with Jupiter and Saturn migrating through the 1:2 mean motion resonance, Marzari and Scholl (2007) showed that a secular resonance (*family III*) between $g$ and one of the two eigenfrequencies of the planetary system sweeps the Trojan region, leading to a chaotic evolution. This instability is reinforced by the sweeping of f*amily II* and *family IV* resonances (Morbidelli et al. 2005, Robutel and Bodossian 2009), with *family IV* resonances being more effective close to the resonance. *Family II* resonances with $\psi_{2,1}, \psi_{3,1}, \psi_{4,1}$ and $\psi_{5,2}$ contribute to instability with different strengths during the migration of Jupiter and Saturn through the 1:2 and towards the 2:5 mean motion resonances.

According to the numerical explorations of Nesvorný and Dones (2002) and Marzari et al. (2003) with chaos detection tools, Saturn Trojans are mostly unstable and any primordial population should have been severely depleted at present. The fast diffusion in the phase space is due to *family III* secular resonances like the $2g_6 - g_5$ and *family II* resonances with $\psi_{5,2} = 5n_J - 2n_S$. The same fate is shared by Uranus Trojans which are affected by secular resonances with $s_7, g_7$ and $g_5$ (Marzari et al. 2003b) and by family II resonances with $\psi_{2,1} = 2n_U - 1n_N$ (Nesvorný and Dones 2002). The situation is different for Neptune Trojans, which, in spite of some perturbations from the $s_8$, have large regions of stability with low diffusion speed (Marzari et al. 2003b, Nesvorný and Dones 2002).

The long term stability of Venus, Earth and Mars Trojans have been investigated mostly with Laskar's FMA (Scholl et al, 2005a, 2005b, Dvorak et al. 2012, Marzari & Scholl 2013). These studies show that Trojans of the terrestrial planets are predominantly perturbed by *family III* secular resonances with the eigenfrequecies $g_2, g_3, g_4, g_5$(V, E, M) and $s_3, s_4$(M). Due to the influence of these resonances, Venus Trojans are unstable with a half-life of about $6 \times 10^8$ years which is further reduced when the Yarkovsky effect is included in the numerical integration of fictitious populations of tadpole orbits. Earth Trojans have, by contrast, large stability regions up to $40^o$deg in inclination, with the peculiarity of favoring middle to large libration amplitude orbits for long term survival, contrary to what is observed for Trojans of the outer planets. The dynamical stability of Mars Trojans is granted only for inclinations between 15$^o$ and 30$^o$, and, even in this case, the Yarkovsky force has some perturbing effect only when very small bodies (in the m-size range) are considered.

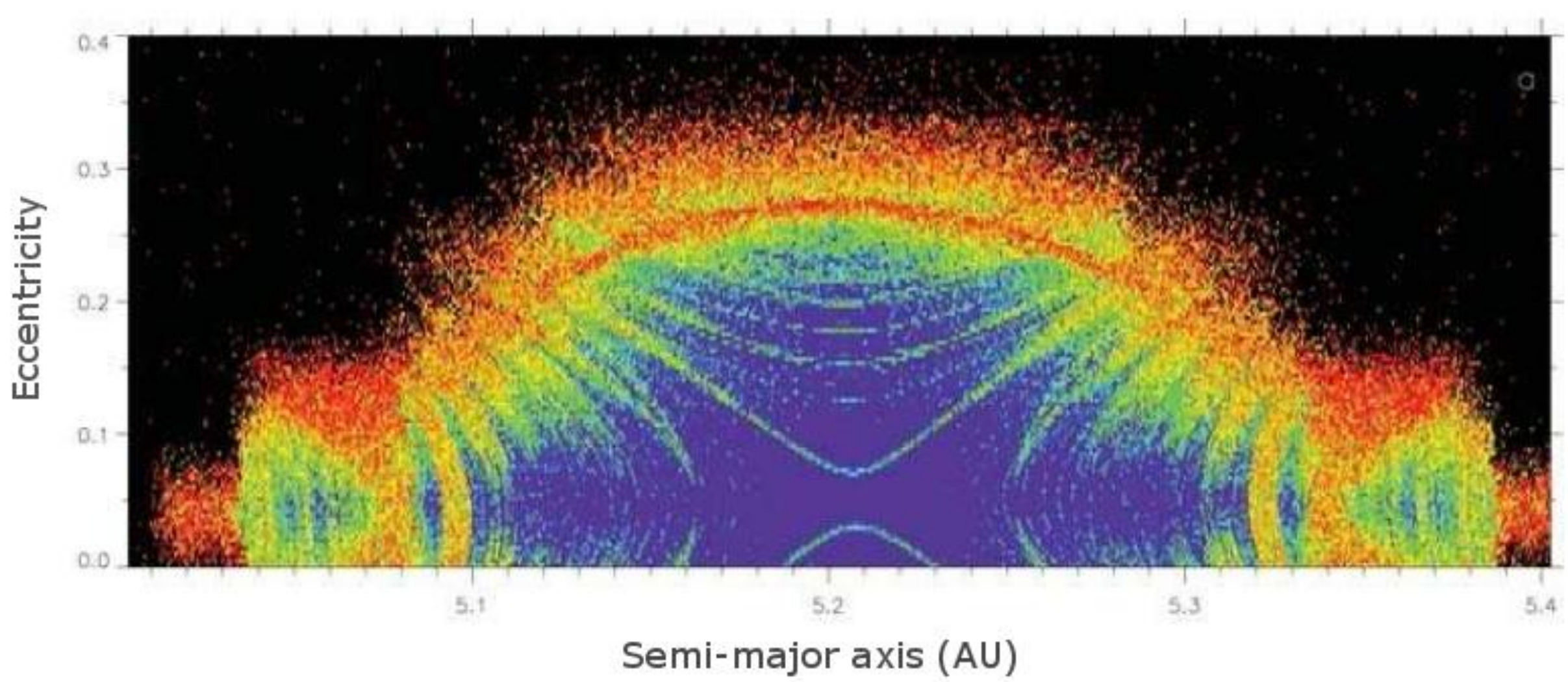

*Fig 9: Diffusion map around L4 for an N-body model including the outer four planets. Blue color indicates stable orbits while red corresponds to highly chaotic motion. The black zone marks trajectories that lead to ejection from L4 on a short timescale. From Robutel & Gabern (2006).*

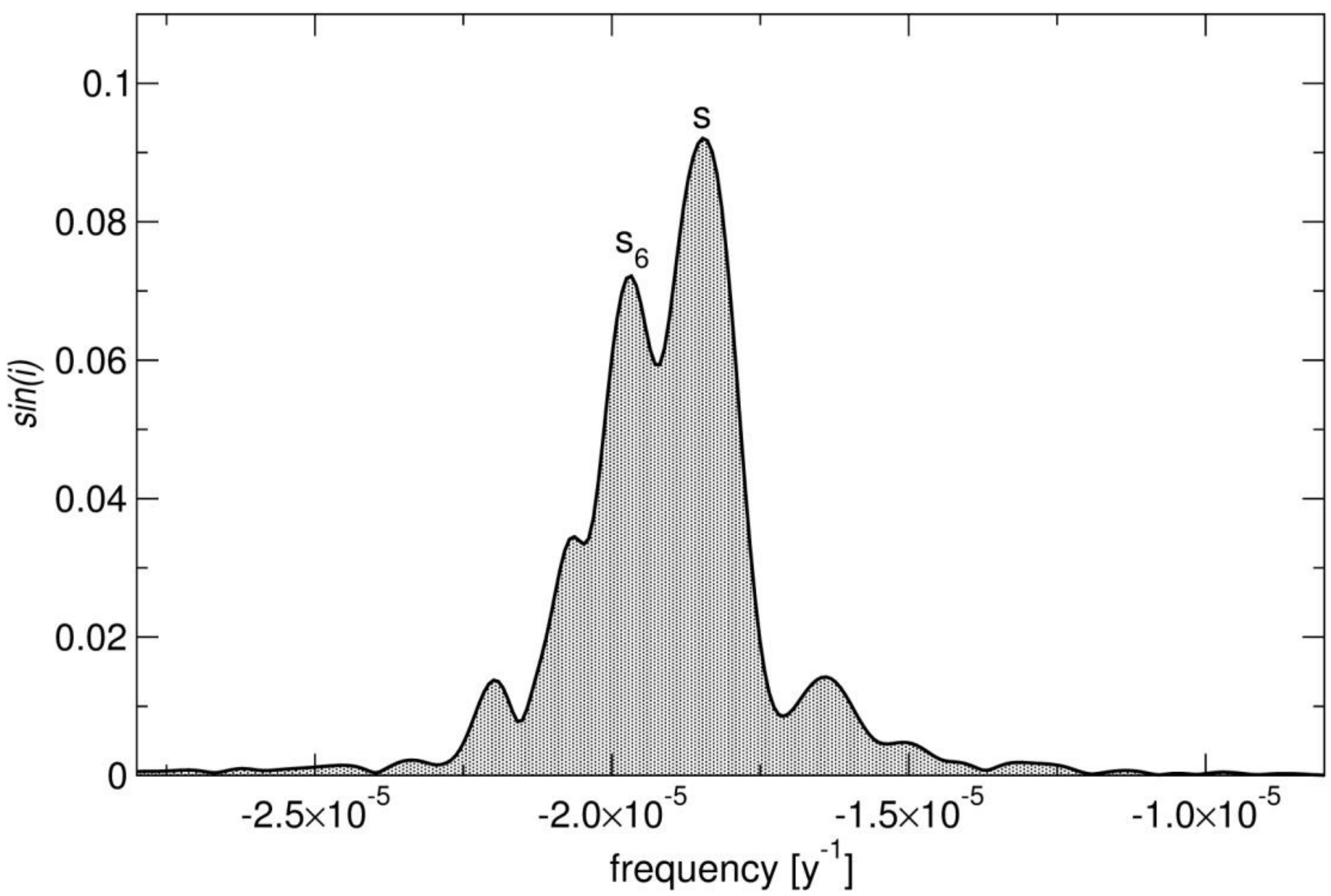

*Fig. 10: Power spectrum of the p,q non-singular variables for a Trojan orbit close to the* $s_6$*secular resonance. The forced component due to* $s_6$*is comparable to the proper frequency* $s$*. From Marzari et al. ( 2003a).*

**3.3. Collisional Evolution**

The collisional evolution of Trojans asteroids, initially explored by Marzari et al. (1996), has been recently revisited by De Elia and Brunini (2007, 2010). The newer work employs a refined collisional model that includes an updated treatment of the fragmentation physics, the escape of bodies suffering impacts that eject them out of the Trojan swarm, and the effects of Poynting-Robertson drag on small particles in the μm-size range.

They find that the size distribution for diameters larger than about 60 km has a power-law slope that is substantially unaltered after 4.5 Gyr of evolution. The measured slope would therefore be primordial, reflecting the size distribution of the planetesimals that were trapped as Trojans during the early evolution of the Solar System. This result is agreement with the suggestion of Morbidelli et al. (2009) that Trojans and KBOs share a common origin not only on a dynamical ground, but also because they have similar slopes to their size distributions, indicating a common origin.

Below 60 km, collisions dominate the evolution, and the slope of the size distribution relaxes towards the Dohnanyi's equilibrium value. The primordial population inferred by the models of De Elia and Brunini (2007, 2010) would include about $1 \times 10^8$ bodies larger than 1 km of which $1 \times 10^6$ would survive at present. The erosion of the Trojan population leads to the formation of families, injection of bodies into Centaur and Jupiter Family Comet (JFC) trajectories (Marzari at el. 1995, Levison et al. 1997), and the formation of a dusty ring around the orbit of Jupiter. According to De Elia and Brunini (2007), the flux of Trojans into the current population of Centaurs and JFCs is negligible with about 50 objects larger than 1 km in diameter per Myr from the $L_4$ swarm. The flux from $L_5$ is expected to be even lower due to the reduced number of Trojans populating the $L_5$ swarm. This estimate is about two orders of magnitude smaller than that given by Marzari et al. (1996) and the difference can mostly be ascribed to the different scaling laws adopted in the collisional models.

A few family forming events are expected for bodies larger than 50 km which would produce the largest families predicted by Beaugè and Roig (2001) and observed by Dotto et. al. (2006) and Fornasier et al. (2007). The robustness of the families identified by Beaugè and Roig (2001) have been recently questioned by Broz and Rozehnal (2011) on the ground that the newly discovered Trojans dilute the clusters, interpreted as families, into the background. They claim that only the large family Eurybates is a real outcome of a breakup event, and the Ennomos group may be real. Family membership is based on proper elements, and Di Sisto et al. (2014) found that computation of proper elements for non-numbered Trojans may not be reliable, even if observations are available over multiple oppositions. Caution is therefore advised in using non-numbered asteroids in the family identification process.

De Elia and Brunini (2010) also explored the production of dust by collisions within the $L_4$ swarm and its lifetime against the Poynting-Robertson drag erosion. They estimate that the present thermal emission in the $L_4$ jovian swarm could be as high as $\sim 3.2 \times 10^{-8} - 3.4 \times 10^{-8}$ $L_{sun}$, comparable to the luminosity of the inner Solar System dust produced by asteroid collisions and cometary activity. This interesting prediction suggests that extrasolar giant planets could be detected also by the emission of a dusty ring produced by collisions of putative Trojans.

## 4. Discussion and Future Directions

The Trojan asteroids remain one of the most fascinating and enigmatic group of small bodies in the Solar System. Perhaps the best established property of the Trojans is their dynamical stability. Although several mechanisms are capable of capturing Trojans, the "Jumping Jupiter" version of the Nice model does the best job of matching the orbital distributions and, at the same time, fits the total population, size distribution and $L_4/L_5$ population asymmetry in a way that is consistent with the overall dynamical evolution of the outer Solar System. As a result, it has become widely accepted that the majority of Trojan asteroids are likely refugees from the primordial Kuiper Belt, and therefore have genetic affinities to the Scattered (or Scattering) and Hot Classical KBOs.

Studies of the physical properties of Trojans, however, do not paint such a clear picture. The low albedos and featureless spectra leave the interpretation of surface compositions open. Direct spectral comparisons with KBOs show significant differences between the two populations. KBOs have a much wider range of albedos than Trojans, extending, in particular, to higher albedos. Whereas the Trojans and small KBOs (and Centaurs) both exhibit color bimodalities, the color groups do not overlap. The "ultra-red" or RR (Barucci et al. 2005) spectral group of KBOs are completely absent from the Trojan swarms. The "red" Trojan group overlaps with the BR (moderately red) spectral group of KBOs and Centaurs, and the "less-red" Trojan group does not have a clear analog among KBOs, though some KBOs do overlap this group spectrally.

The possibility that surface compositions may have been modified by the changing irradiation and thermal environments as KBOs migrated inward is intriguing. The presence of ultra-red slopes, strong, broad absorptions at 3.6 and 4.5 μm, and a feature near 2.35 μm attributed to methanol suggest the presence of complex organics on at least some KBOs and Centaurs (e.g., Cruikshank et al. 1998, Barucci et al. 2006, Emery et al. 2007**,** Dalle Ore et al. 2013**,** Dalle Ore et al. 2015). Irradiation of these materials could lead to a decrease in spectral slope (e.g., Moroz et al. 1998). Nevertheless, the absence of any organic absorptions, combined with clear signatures of silicate dust on Trojans, challenges any simple irradiation hypothesis (e.g., Jewitt 2000, Melita et al. 2009). Periods of cometary activity are likely involved (Melita and Licandro 2012), especially if, as Morbidelli et al. (2005) discuss, the KBOs that migrated into the Trojan swarms spent significant time in orbits that brought them close to the Sun, but no satisfactory surface evolution scenario has yet been put forth that could explain the two Trojan spectral groups from a single parent population.

Guilbert-Lepoutre (2014) demonstrated that $H_2O$ ice can be stable on Trojans for the lifetime of the Solar System if covered by ~10 m of dust at the equator (~10 cm near the poles). We might therefore expect to see some evidence for exposure by impact, particularly for smaller Trojans. Nevertheless, there is no indication that albedos increase even for the smallest Trojans currently observable. On the other hand, the very weak 3-μm absorption reported by Brown (2014) may be the first hint of a subsurface ice reservoir. Whether Trojans formed near 5AU or in the Kuiper Belt, they should have accreted abundant $H_2O$ ice, so the absence of any spectral signature of ice among the Trojans is curious.

Very little is known about the interior structure of Trojan asteroids. The only direct constraint comes from the densities that have been derived for a few Trojans. The low density of Patroclus suggests a porous, ice-rich interior. Conflicting reports regarding Hektor's density make it difficult to assess the potential interior structure. Potential geochemical evolution of the interiors is also an open question. The apparently comet-like silicate dust reported by Emery et al. (2006) may suggest very little parent-body processing. On the other hand, the model of a salt-rich surface described by Yang et al. (2012) would require a significant amount of thermal processing of original primitive ices and silicates. Without detailed models of the thermal and chemical evolution of Trojan asteroids, it is difficult to interpret present surface compositions in terms of possible interior structures and evolutionary scenarios.

The two primary questions about the Trojans remain 1) where did they form? and 2) what are they made of? As described above, the past decade has seen advances in dynamical simulations and observations of physical properties that have drawn us significantly closer to the answers to these questions, but several particular areas are ripe for future investigations.

- What is the size distribution at small sizes?
    - How different is it for the two spectral groups?
    - What is the significance of the difference between $L_4$ and $L_5$ at small sizes?
- Why is the one robust family (Eurybates) spectrally anomalous?

- Does the capture mechanism preferentially select any original population?
    - What fraction, if any, might be from the Jupiter region (or closer)?
- Is the dust environment as predicted?
- How is Kuiper Belt material (simple ices, organics, cometary silicates) modified as a body migrates inward to the thermal and irradiation environment of the Trojans?
- What is the ice fraction in the interiors of Trojans?
    - How deep is any extant ice buried?
    - Are any ices aside from $H_2O$ present?
- What is the nature of the low albedo material on Trojan surfaces?
    - Are organic materials present on the surfaces? If so, what is its structural and chemical form?
- Are smaller Trojans different in spectra and/or albedo from their larger siblings?
- Are Trojan silicates (and ices, if present) more similar to cometary or asteroidal material?
- Geologically, do Trojans resemble asteroids, comets, or irregular satellites (i.e., Pheobe)?
- Is there any outgassing or other source of extended emission on Trojans?
- Is the surface structure extremely porous ("fluffy"), or are the fine-grained silicates embedded in some matrix, such as a salt?
- What are the range of possible thermal and chemical histories for the interiors of Trojans, and how can current or future observations constrain those possible histories?

Fortunately, exciting prospects are on the horizon for learning more about Trojan asteroids and answering some of these important questions. In the nearest term, recent work by, e.g., Marssett et al. (2014), Brown (2014), and Wong et al. (2014) demonstrate the possibility of pushing spectral studies to smaller sizes with existing telescopes and instrumentation. Such observations will continue to test the hypothesis that the interiors of Trojans are distinct from their surfaces and, if the Brown (2014) and Wong et al. (2014) studies are indicative, may well surprise. Rotation properties of Trojans have long been uncertain, and the new works revealing such flat period distributions are important for understanding non-gravitational torques at large distances from the Sun. Additional information on amplitudes and spin-pole orientations will provide important constraints on these torques as well as the collisional environment. Extending spectral studies to the UV, which is currently possible with the Hubble Space Telescope, would also provide new insight into the silicates (and any ices) on the surfaces.

Through a series of workshops focused on "*In Situ* Science and Instrumentation for Primitive Bodies" that included scientists of diverse backgrounds (observers, laboratory cosmochemists, dynamical modelers), supported by the Keck Institute for Space Studies, Blacksberg et al. (2013) concluded that devising advances in instrumentation for surface science on primitive bodies, particularly the Trojan asteroids, is premature because of the fundamental uncertainties that remain in what chemical, mineralogical, and isotopic compositions to expect on the surfaces. They instead recommended a program of laboratory study to investigate the potential alteration pathways that KBO surface materials might go through on their journey to the Trojan swarms (or closer). Such laboratory work has the potential to enable direct tests of potential dynamical pathways by linking them to expected compositional changes and would thereby lead to significant advances in understanding Trojan surfaces even from the current observational dataset.

Several planned and potential ground-based and space-based survey programs are expected to lead to significant improvements in discovery and characterization of Trojans. ESA's Gaia mission, which started its science observations in mid-2014, will provide spectral characterization at visible wavelengths of asteroids down to V~20. The Large Synoptic Survey Telescope (LSST) and the Panoramic Survey Telescope and Rapid Response System (Pan-STARRS), when they are fully operational are anticipated to discover and record colors of hundreds of thousands of Trojan asteroids. Significant interest has been expressed recently in an infrared space telescope for asteroid discovery, for hazard mitigation, to support human exploration, and for science. The benefit of infrared discovery, particularly with at least two infrared pass-bands, as with the NEOWISE survey (Mainzer et al. 2011), is that the discovery data directly provide sizes. Such a mission would likely discover, and measure sizes of, large number of Trojans, and it is also likely that temporal coverage would enable estimates of thermophysical properties of the surfaces.

In terms of surface characterization, the next leap forward in terms of Trojan asteroids will probably come from the James Webb Space Telescope (JWST). With spectral coverage from 1 to 28 μm, JWST is ideally suited for searches for ices and organics, characterization of silicates, and determination of thermophysical properties. The sensitivities of JWST at wavelengths longward of 2.5 μm will significantly exceed those of current ground-based telescopes, and will enable observations of much smaller Trojans than is now possible.

Perhaps the most exciting future prospect for advances in our understanding of Trojans is the possibility of spacecraft missions to observe this population close-up. Because Trojans are, in many ways, key to understanding the evolution of the Solar System, there has been significant international interest in targeting the Trojan asteroids with an upcoming mission. The Trojans have been called out in each of the two last decadal surveys for NASA planetary science, making the short list for desired missions in the New Frontiers class. Lamy et al. (2012) make the case for a mission to the Trojans in the context of ESA's program of Solar System exploration. Diniega et al. (2013) report the results of a JPL Planetary Science Summer School design exercise for a mission to the Trojans. The Japanese space agency, JAXA, has been developing solar sail technology that may be well-suited to providing propulsion for a deep-space mission to the Trojan asteroids (e.g., Yano et al. 2013).

The details of a spacecraft mission to the Trojans could take many forms. A mission that includes flybys of several objects would provide important information on diversity among the Trojan swarms. An orbiter mission could return detailed geologic and spectral maps, information on the interior structure, and chemical composition of the body. A coordinated orbiter and lander could provide even more detailed "ground-truth" for the orbital investigation. The key is to get close enough to one, or better yet several, Trojans to reveal them as geologic bodies rather than the point sources we know them as from Earth.

There is growing momentum behind determining the nature of this large, enigmatic population of primitive bodies, and it would be reasonable to expect, by whatever avenue the information may come, a revolution in the understanding of Trojan asteroids by the time *Asteroids V* goes to press.